# Effect of Longitudinally Varying Cloud Coverage on Visible Wavelength Reflected-Light Exoplanet Phase Curves

Matthew W. Webber[1], Nikole K. Lewis[1,5], Mark Marley[2], Caroline Morley[3], Jonathan Fortney[3], Kerri Cahoy[1,4]

[1] Department of Earth, Atmospheric, and Planetary Sciences. Massachusetts Institute of Technology (MIT) Cambridge, MA
[2] NASA Ames Research Center, Moffett Field, CA
[3] Department of Astronomy & Astrophysics, University of California, Santa Cruz, CA 95064, USA
[4] Department of Aeronautics and Astronautics. Massachusetts Institute of Technology (MIT) Cambridge, MA
[5] Sagan Postdoctoral Fellow

**ABSTRACT**

We use a planetary albedo model to investigate variations in visible wavelength phase curves of exoplanets. Thermal and cloud properties for these exoplanets are derived using one-dimensional radiative-convective and cloud simulations. The presence of clouds on these exoplanets significantly alters their planetary albedo spectra. We confirm that non-uniform cloud coverage on the dayside of tidally locked exoplanets will manifest as changes to the magnitude and shift of the phase curve. In this work, we first investigate a test case of our model using a Jupiter-like planet, at temperatures consistent to 2.0 AU insolation from a solar type star, to consider the effect of $H_2O$ clouds. We then extend our application of the model to the exoplanet Kepler-7b and consider the effect of varying cloud species, sedimentation efficiency, particle size, and cloud altitude. We show that, depending on the observational filter, the largest possible shift of the phase curve maximum will be ~2-10° for a Jupiter-like planet, and up to ~30° (~0.08 in fractional orbital phase) for hot-Jupiter exoplanets at visible wavelengths as a function of dayside cloud distribution with a uniformly averaged thermal profile. The models presented in this work can be adapted for a variety of planetary cases at visible wavelengths to include variations in planet-star separation, gravity, metallicity, and source-observer geometry. Finally, we tailor our model for comparison with, and confirmation of, the recent optical phase-curve observations of Kepler-7b with the *Kepler* space telescope. The average planetary albedo can vary between 0.1 – 0.6 for the 1300 cloud scenarios that were compared to the observations. Many of these cases cannot produce a high enough albedo to match the observations. We observe that smaller particle size and increasing cloud altitude have a strong effect on increasing albedo. In particular, we show that a set of models where Kepler-7b has roughly half of its dayside covered in small-particle clouds high in the atmosphere, made of bright minerals like $MgSiO_3$ and $Mg_2SiO_4$, provide the best fits to the observed offset and magnitude of the phase-curve, whereas Fe clouds are found to have too dark to fit the observations.

1. INTRODUCTION

Observations of exoplanet systems at both infrared and visible wavelengths have revealed the presence of variations in the system flux that can be linked to the orbital phase of the planet (e.g. Knutson et al. 2007, Harrington et al. 2006, Snellen et al. 2009, Demory et al. 2011). Most of the exoplanet systems that show phase variations are host to Jupiter-sized planets that orbit within 0.1 AU of their host stars, so called 'hot Jupiters'. Due to their close-in orbits, hot Jupiters are generally assumed to be tidally locked to their host stars much like the moon is tidally locked to the Earth. If heat is not transported efficiently away from the substellar point, the peak thermal emission from a tidally-locked exoplanet on a circular orbit should occur when the substellar point aligns directly with the observer (in this work, we designate this alignment as phase angle $\alpha = 0°$). For exoplanets that transit their host star as seen from Earth, so called transiting exoplanets, this maximum emission would align with the secondary eclipse (Barman et al. 2005). The exoplanet infrared phase curves observed to date have shown a great diversity in both the amplitude of the phase variations and the location of the peak in the planetary flux with respect to orbital phase (e.g., Harrington et al. 2006, Knutson et al. 2007, Cowan et al. 2007, Crossfield et al. 2010, Cowan et al. 2012, Knutson et al. 2012, Lewis et al. 2013, Maxted et al. 2013, Zellem et al. 2014). At thermal wavelengths, a shift away from $\alpha = 0°$ in the phase curve peak of the planetary flux can be a sign of asynchronous rotation. If the planet is gaseous, circulation and asynchronous rotation correspond to a similar situation. A shift would also occur for an asynchronous planet with no atmosphere, due to the thermal inertia of the surface (e.g. Selsis et al. 2013, Samuel et al.

2014). However, if synchronous rotation due to tidal locking is assumed, shifts at infrared wavelengths can be attributed to the presence of winds circulating heat away from the substellar point (e.g., Knutson et al. 2007, Rauscher et al. 2008, Showman et al. 2009).

At visible wavelengths, thermal emission contributes minimally to observed phase variations because the peak of the blackbody spectrum of a hot Jupiter is generally at infrared wavelengths. Instead, the planet will contribute mostly reflected light to the phase curve in the visible spectrum (Cowan et al. 2007). For such reflected-light phase curves, the flux maximum is similarly expected to be at $\alpha = 0°$ when the full disk is illuminated for the observer and expected to decrease when a smaller crescent of the planet is visible (Madhusudhan et al. 2012). To date, visible-wavelength phase-curve observations have predominantly shown the maximum in the planetary flux to be well aligned with $\alpha = 0°$, or secondary eclipse, for transiting exoplanets (e.g. Rowe et al. 2008, Snellen et al. 2009, Borucki et al. 2009, Berdyugina et al. 2011, Esteves et al. 2013). However, Demory et al. (2013) recently observed an offset in the peak of the planetary flux away from secondary eclipse at visible wavelengths for Kepler-7b, and discuss how non-uniform albedo due to clouds could explain this observation; which we will further investigate in this manuscript. Esteves et al. (2014) also presents a large sample of optical phase curves with diverse peak shifts and behaviors, including Kepler-7b.

There are several factors that can contribute to variations in visible wavelength phase curves. Arnold and Schneider (2004) show that the presence of planetary rings can shift the phase curve maximum. For rocky planets, the non-uniform albedo of land/water configurations may help shape the reflected light phase curve (Williams and Gaidos 2008, Mallama 2008). Ellipsoidal variations (Welsh et al. 2010) or an oblate planet with a non-edge-on inclination (Dyudina et al. 2005) should also be considered when evaluating contributions from physical processes that shape optical phase curves. In this work, we consider the contribution of non-uniform cloud coverage to the visible wavelength phase curve amplitude and offset for hot Jupiters.

We use an albedo spectra model to show that the shape and maximum of the reflected-light phase curve is dependent on the spatial distribution of cloud coverage on an exoplanet. In the case of tidally locked planets, there will likely be a large temperature contrast between the dayside and nightside of the planet, which will lead to large gradients in the composition, average particle size, and altitude of any clouds that may form (Morley et al. 2012, Parmentier et al. 2013). Infrared phase curve observations, combined with atmospheric models, have shown that close-in tidally-locked hot Jupiters develop rapid eastward equatorial jets (Showman et al. 2009, Showman & Polvani 2011) that will cause shifts in both thermal and cloud patterns, and thereby alter the overall optical albedo spectrum of the planet.

The albedo model used in this work was developed by McKay et al. (1989) and modified by Marley and McKay (1999), Marley et al. (1999), and Cahoy et al. (2010) to study the atmospheres of solar system planets and extrasolar giant planets. Here we consider the albedo spectra of both a hypothetical solar system Jupiter, at temperatures similar to 2.0 AU insolation from a solar type star, and the hot Jupiter Kepler-7b (Latham et al. 2010). The former case allows us to compare to the results of Cahoy et al. (2010) who studied a similar planet; while the latter case permits the study of non-uniform coverage of forsterite ($Mg_2SiO_4$), enstatite ($MgSiO_3$), and iron (Fe) clouds on the albedo spectrum as a function of orbital phase. For each planet, an average temperature-pressure profile is used for all cases and location; there is no thermal gradient. Therefore, all longitudinal albedo variation is due only to prescribed cloud coverage. In the discussion section, we use our albedo and cloud model results to support the interpretation of the published visible wavelength phase curve for Kepler-7b from Demory et al. (2013).

## 2. APPROACH

To model the thermal and chemical structure of the atmospheres of exoplanets we use a code that has been widely applied to the atmospheres of solar system planets (Mckay et al. 1989, Marley &

Mckay 1999), brown dwarfs (Marley et al. 1996, 2010), and exoplanets (Fortney et al. 2005, 2008). The atmosphere model is one dimensional (1D), plane-parallel, and converges to radiative-convective equilibrium temperature structures. Atmospheric opacities (see Freedman et al. 2008, 2014) are tabulated using the correlated-k method (Goody et al. 1989). Chemical equilibrium abundances (Lodders & Fegley, 2002, 2006; Lodders 2009) are calculated at solar abundances (Lodders 2003).

First, the 1D radiative-convective model computes atmospheric profiles for Jupiter at 2.0 AU and Kepler-7b. The temperature-pressure profiles and cloud profiles from the radiative-convective solutions are used by the albedo model to produce albedo spectra at visible wavelengths based on parameters that include planet-star separation, gravity, metallicity, and source-observer geometry. The albedo model inputs a one dimensional 70-layer temperature-pressure profile for Jupiter, and 60-layer temperature-pressure profile for Kepler-7b (Figure 1). For detailed description of the radiative-transfer and scattering methods used in the albedo spectra model see Cahoy et al. (2010), which shows the phase-dependence of Jupiter and Neptune-like models with varying planet-star separation and metallicity. Cahoy et al. also make coarse-resolution spectra and color-color predictions for direct imaging of exoplanets.

Our one-dimensional cloud model was developed in Ackerman & Marley (2001), and treats the balance of sedimentation of particles against updrafts of particles and condensable vapor. A key parameter of the model is $f_{sed}$, the sedimentation efficiency parameter, which is often adjusted to best fit observations. Large values of $f_{sed}$ correspond to efficient sedimentation, large cloud particles, and vertically compact clouds with relatively small vertical extents and optical depths. Small values of $f_{sed}$ lead to small particles, and vertically taller clouds with larger optical depths. Cloud optical depths are calculated using Mie theory. For L-type brown dwarfs dominated by silicate and iron clouds, $f_{sed}$ values from 1-3 are typically found (Saumon & Marley, 2008; Marley et al. 2010), while for the cooler T-type dwarfs where sulfide and salt clouds may be present, $f_{sed} \sim 5$ appears to best fit observations (Morley et al. 2012, Leggett et al. 2013). At conditions relevant for the atmospheres of hot Jupiters, there are not yet strong constraints on $f_{sed}$. The evidence for high clouds of very small particles in the transmission spectrum of HD 189733b (Evans et al. 2013, Sing et al. 2011, Pont et al. 2008, Lecavelier des Etangs et al. 2008) may suggest much lower values of $f_{sed}$ for hot Jupiters. The observations of super-Earth GJ 1214b were also matched with a low $f_{sed} = 0.1$ model with 50x solar-abundance atmosphere. (Morley et al. 2013).

The albedo model also uses the cloud opacity profiles, which vary with pressure and wavelength, from the 1D radiative-convective model for both cases. Clouds are predicted based on the condensation curves of the species compared to the pressure profile of the atmosphere ($H_2O$ or $NH_3$ curves for Jupiter-like models and $MgSiO_3$, $Mg_2SiO_4$, and Fe curves for Kepler-7b). Albedo spectra can be computed with or without clouds. Figure 2 shows the case where half the dayside is cloud-covered (red) and half the dayside is cloudless (blue). This is the 90° offset case; clouds are offset by 90° into the dayside. As the phase angle changes (i.e. the planet orbits), the incident and reflected light angles change. In Figure 2 the observer does not see the shaded-out portions of the disk at these wavelengths. Smaller sections of the dayside are reflecting light toward the observer as the phase angle moves away from zero.

In the context of Figure 2, each point on the disk relates an individual albedo spectrum to corresponding weights for a Tchebyshev-Gauss integration. The ability to place individual spectra that are unique to each location allows us to simulate non-uniform cloud coverage. Tchebyshev-Gauss quadrature was used to integrate over the planetary disk with 10 points in the Tchebyshev dimension (ten points in latitude south to north) and 100 points in the Gaussian dimension across longitude (Horak 1950, Horak and Little 1965, Davis and Rabinowitz 1956). A spectrum is computed for each point depending on the specific cloud coverage and source-observer geometry for that location.

The Gaussian Angles, $\theta_{Gi}$, and Gaussian weights, $W_{Gi}$, are based on the i$^{th}$ root of the G$^{th}$ Legendre Polynomials. The Tchebyshev Angle $\theta_{Ti}$, and Tchebyshev weights, $W_{Ti}$, are found from:

$$\theta_{T_i} = \cos\left(\frac{i\pi}{T+1}\right)$$

$$W_{T_i} = \frac{\pi}{T+1}\sin^2\left(\frac{i\pi}{T+1}\right)$$

The angle of incidence, $\mu_0$, and emergent angle, $\mu_1$, from each atmospheric section and phase angle, $\alpha$, are calculated from:

$$\nu = \frac{1}{2}\left(\theta_G - \frac{\cos\alpha - 1}{\cos\alpha + 1}\right)[\cos\alpha + 1]$$

$$\mu_0 = \sin[\cos^{-1}(\theta_T)]\cos(\nu - \alpha)$$

$$\mu_1 = \sin[\cos^{-1}(\theta_T)]\cos(\nu)$$

The resulting albedo spectrum, $A(\lambda, \alpha)$, as a function of phase is derived from the sum of each point's intensity, $I_{ij}(\lambda,\alpha)$, with weight based on the position on the planetary disk:

$$A(\lambda, \alpha) = \frac{1}{2}(\cos\alpha + 1)\sum_{i=1}^{G}\sum_{j=1}^{T} W_{G_i} W_{T_j} I_{ij}(\lambda, \alpha)$$

For $\alpha = 0°$, $A$ is the geometric albedo (i.e. relative to a Lambertian disc). The spectra at all subsequent phase angles represent the albedo relative to the $\alpha = 0°$ Lambertian disc. By integrating the albedo spectra over a desired wavelength bandpass at each phase angle, we can find the relative reflected-light intensity for an observer as a function of phase. For this analysis, we use a general "top-hat" bandpass from 350-850 nm. The maxima of the phase curves are derived from fitting a sine function to a 60° range of points around the phase-curve peak. Error-bars are based on the 95% confidence intervals of the fit parameters. In this study we nominally assume that clouds either 1) form on the nightside of the planet and are subsequently 'shifted' onto the dayside of the planet across the western terminator due to winds, or 2) form due to an asymmetrical thermal distribution of the dayside such as the one measured for HD189733b by Knutson et al. (2007). We first assume a 30° cloud offset as representative of the expected offset from thermal observations of Hot Jupiters (e.g. Knutson et al.). We then use an offset of 90° to show the maximum shift due to this effect.

## 3. RESULTS & DISCUSSION

Theoretical phase curves are presented in this section for different planet models where each is normalized to the curve with the highest albedo for comparison (relative albedo). While only the 30° and 90° cloud offset cases are shown for each planet, our model can compute the phase curve for the range of clouds offset onto the dayside between 0° and 180°. For clarity, we use the term "offset" to refer to the degree of cloud coverage on the dayside of the planet with respect to the western terminator, and the term "shift" to refer to the deviation of the phase curve peak vs. phase angle. As expected, the non-uniform cloud coverage causes an asymmetry of the phase curve and a shift in the peak brightness. The phase curves will increase in brightness as more clouds are added to the dayside (e.g. the 30° cloud offset will be have a higher relative albedo than the 90° cloud offset). Table 1 presents the shift of phase curve maxima for each cloud offset. The scale of these shifts is dependent on the unique spectra of each planet.

### 3.1 JUPITER 2.0 AU

The test case for this model is the temperature-pressure profile of Jupiter at 2.0 AU from a solar-type star (Cahoy et al. 2010). We note that a Jupiter-like planet at 2.0 AU from a solar-type star will not be tidally locked and will not exhibit the same strong day-night temperature gradient seen in hot-Jupiter atmospheres. In fact, Karkoschka (1992) observed no longitudinal or diurnal asymmetry for Jupiter in its true orbit. We simply use the Jupiter at 2.0 AU test case to benchmark against the Cahoy et al. (2010) albedo models and to explore the albedo spectra of water that may form in the atmospheres of planets close-in to less luminous M and K-dwarfs (c.f., Kasting et al. 1993). Cloudy and cloudless albedo spectra were computed for each of the 1000 points in the Tchebyshev-Gauss grid. The resulting spectra for a 90° offset, integrated over the planetary disk and scaled by the Tchebyshev-Gauss weights, are shown in Figure 3 for a range of phase angles. Figure 3 is used to illustrate the spectral difference that lead to changes in the phase curve. Note, several intermediate spectra ($\alpha = \sim 6 - 30\ °$) increase above the $\alpha = 0°$ spectrum after ~550 nm, causing the shift in the phase curve (there is a similar effect in Figure 5). Figure 4 shows four phase curves for Jupiter at 2.0 AU: a) the dayside fully covered with clouds, b) the dayside half covered with clouds (90° offset), c) 30° offset of cloud coverage into the dayside, and d) a cloudless dayside. The shift of the phase curve maximum past eclipse (and away from the uniform cases) for the 30° cloud offset is 2.0 ± 0.4° and for the 90° offset is 10.1 ± 0.8°.

### 3.2. KEPLER 7b

Our approach is repeated using a theoretical temperature-pressure profile for Kepler-7b (Demory et al. 2013). Kipping & Bakos (2011) and Demory et al. (2011) concluded a high average geometric albedo (~ 0.3) from secondary eclipse observations, at visible wavelengths, for Kepler-7b possibly due to Rayleigh scattering combined with clouds or haze, but could not entirely rule out the possibility that the deep eclipse was due to thermal emission. Secondary eclipse observations of the Kepler-7 system at infrared wavelengths with the *Spitzer* space telescope show that Kepler-7b exhibits very little thermal emission, further strengthening the case for optically thick clouds in Kepler-7b's atmosphere (Demory et al. 2013). Given the theoretical temperature structure expected for Kepler-7b, a wide array of equilibrium cloud species are expected to form near photospheric pressures (see Morley et al. 2012 for condensation curves). As discussed previously, a strong eastward equatorial jet is expected to form in the atmospheres of hot-Jupiters like Kepler-7b (e.g., Showman et al. 2009). Although a thermal phase-curve cannot be measured for Kepler-7b, given the recent *Spitzer* observations to confirm the presence and strength of this equatorial jet, the same predicted wind and thermal patterns may lead to clouds whose properties vary strongly with longitude on the dayside. Here, we consider clouds with a log-normal particle size distribution with a mode of 0.5 μm (see Demory et al. 2013). Figure 6 again shows four phase curves (for Kepler-7b): a) the dayside fully covered with clouds b) the dayside half covered with clouds (90° offset) c) 30° offset of cloud coverage into the dayside and d) a cloudless dayside. For the 30° cloud offset, the phase peak shifted 9.8 ± 1.7°. For the 90° cloud offset example, the peak shift was 17.8 ± 2.0°.

For observations of a transiting exoplanet with a uniform disk, the phase curve maximum and secondary eclipse will occur at a phase angle of 0°. When there is non-uniform cloud coverage, the maximum of the phase curve will shift with increasing amounts of longitudinal varying clouds. From the observed phase shift estimates of the cloud properties such as coverage, average particle size, and potential composition, can be made. These cloud properties can then potentially be linked to the global scale wind and thermal patterns required for their formation.

### 3.3 Non-uniform Cloud Coverage

In each case, as summarized in Table 1, the phase curve maximum increases with the increasing cloud coverage offset. Also, for the 90° cloud offset cases, we see that the phase curves agree with the full cloudy or cloudless models before and after the phase angles ±90°. This is also true but less obvious at -30° and +150° for the 30° offset cases. This overlap of the phase curves agrees with the source-observer geometries where the partial cloudy/cloudless regions would no longer be visible.

For thermal profiles and cloud properties considered in this study, the most reflective case is the completely cloudy planet and the least reflective case is the cloudless planet (Marley et al. 1999, Sudarsky et al. 2000, and Cahoy et al. 2010). Sudarsky et al. (2005) notes that the maximum and shape of the phase curve will inform the presence of clouds. As the distribution of dayside cloud coverage changes, the total relative brightness of the planet increases as well as the shift of the phase curve maximum. The shift of the phase curve maximum will be largest for the 90° offset cases (i.e. the half-and-half hemisphere case). For cloud offsets more than 90°, the shift of the phase curve maximum will decrease and the phase curve will look more and more like completely cloud covered cases (red curves in Figures 3 and 5). Note that if the cloudy hemisphere is instead assumed to be east of the substellar longitude, the shift of the maximum of the phase curve would be negative and occur before secondary eclipse.

The largest possible shifts of the phase-curve maxima for the two atmospheric models considered here are shown in Table 1. The Kepler-7b model showed larger shifts (9.8 ± 1.7° and 17.8 ± 2.0° for 30° and 90° offsets respectively) while the Jupiter 2.0 AU test case showed smaller shifts (2.0 ± 0.4° and 10.1 ± 0.8° for 30° and 90° offsets respectively). The scale of the shift is dependent on the difference between a cloudy and cloudless spectrum as well as the optical properties of the cloud species considered in each case. Also, the slope of the phase curve is steeper for the negative phase angles where the crescent of reflected light is losing (or gaining) bright cloudy area as opposed to the shallower slope of the positive phase angles; in this regime, the changing contribution is from the cloudless area, which is dimmer at visible wavelengths and has less effect as it moves out-of-view.

Figure 7 shows the spectra of cloudy vs. cloudless planets at $\alpha = 0°$. The clouds for Jupiter at 2.0 AU are $H_2O$ condensates. The model used here for hot Jupiter Kepler-7b was previously presented in Demory et al. (2013) and includes the opacity of forsterite ($Mg_2SiO_4$), a representative silicate. Recall in Figure 1 that the condensation curve of forsterite crosses the temperature-pressure profile of Kepler-7b and defines the altitude of the cloud layers. In order to increase the planet's albedo, given the observations of Demory et al. (2013), $f_{sed} = 0.1$ was used in this particular model to compute a number density and particle size distribution for the clouds structure. The clouds were forced to have a particular size distribution, but otherwise the cloud properties were allowed to develop consistently with the thermal profile. This method does not include mass conservation of the prescribed atmospheric abundances but is instructive in demonstrating the effect of high, thick clouds. These small particles are appropriate for the regions of strong updraft seen in the general circulation models (GCM) (Parmentier et al. 2013). For the Jupiter at 2.0 AU model $f_{sed} = 6$ was used. For each case, when clouds are added, the *difference* in reflectivity of the planet roughly increases with wavelength. The effect of Rayleigh scattering is washed out by the onset of clouds with higher albedo in the red wavelengths. The Jupiter case has a slower fall-off for the cloudless spectra so the addition of clouds results in a smaller shift (compared to the Kepler-7b case). Figure 7 show that contributions from cloudy vs. clear spectra to phase curve measurements depend on the particular planet as well as factors such as cloud coverage and observation filters.

### 3.4 Comparison to Kepler-7b Observations

To build upon the work of Demory et al. (2013) who presented a single reasonably well-fitting model, we perform a parameter exploration of cloud composition, sedimentation efficiency, cloud-base altitude, and cloud longitude offset. We can use our results to compute a planet-star flux ratio vs. phase angle for a range of cloud offsets and compare with the observed visible wavelength phase curve of Kepler-7b presented in Demory et al (2013). A model stellar spectrum of Kepler-7, using a NextGen PHOENIX stellar atmospheres model (Hauschildt et al. 1999), is weighted to the Kepler transmission function and integrated to compute an observed stellar flux. Our model stellar spectrum is interpolated from the grid of similar NextGen models[†] to match a set of assumed parameters. For the stellar parameters of Kepler-7 we assume: M/H = 0.0, effective temperature = 5933 K, mass = 1.36 $M_{solar}$, radius = 2.02 $R_{solar}$, surface gravity $\log(g_*)$ [cgs] = 3.96 (Demory et al. 2011). To compute the planet flux, the stellar spectrum is also weighted by the planetary albedo spectrum at each phase angle. This spectrum can be integrated and multiplied by the square of the ratio of the planet radius ($R_p$ = 1.63 $R_J$) to the semi-major axis (a = 0.06246 AU) to find the planet flux.

---

[†] http://phoenix.ens-lyon.fr/Grids/NextGen/

In addition to varying the dayside cloud offset longitude (in 15° increments), we computed a large parameter space of cloud cases to test against the Kepler-7b data set by varying the cloud species, $f_{sed}$, and cloud altitude. For simplicity in comparing this grid of models, a single pressure-temperature profile (Figure 1) is used to determine the cloud structure for each case without iterating a new profile in radiative-convective equilibrium. The average planetary albedo can vary between 0.1 – 0.6 depending on the cloud conditions. An automated pipeline was created to compute albedo spectra vs. phase angle for each case at a range of longitudinal cloud offsets and test their fit to the Demory et al. (2013) phase curve. Overall, 1300 cloud conditions were investigated. The four cloud species used were: $MgSiO_3$, $Mg_2SiO_4$, Fe, and a mixed $Mg_2SiO_4$ + Fe. Figure 8 shows several cases (at α = 0°) with varying species where $f_{sed}$ is held constant. For this condition, $MgSiO_3$ shows the most reflectivity. Each cloud case was computed with $f_{sed}$ = 0.03, 0.1, 0.3, 1.0, and 3.0. Recall, the sedimentation efficiency is used in the cloud model to balance upward turbulent mixing with downward condensation:

$$-K_{zz}\frac{\partial q_t}{\partial z} - f_{sed}w_* q_c = 0$$

Where $K_{zz}$ = eddy diffusion coefficient, $q_t$ = total mixing ratio, $q_c$ = condensate mixing ratio, $w_*$ = convective velocity scale. A low $f_{sed}$ suppresses sedimentation and a higher values lead to thinner, less optically thick, cloud layers. As $f_{sed}$ increases, particle size increases. Figure 9 shows the effect of varying sedimentation efficiency for a fixed cloud species. For Figure 9, the effective particle sizes at the layer with the maximum condensate concentration for each case are about 5, 15, 40, 100, and 220 μm for $f_{sed}$ = 0.03, 0.1, 0.3, 1.0, 3.0 respectively. With large $f_{sed}$ values, or large particles, the planet will appear dark. Sufficiently bright albedos are produced only by using a small $f_{sed}$ (≤ 0.1) in the cloud formation model.

In order to vary the cloud altitude, we shifted the temperature-pressure profile by ±100 K and ±200 K. Figure 10 shows the resulting albedo spectra for a fixed cloud species and sedimentation efficiency. The +200K spectrum is identical to the cloudless spectra because the pressure-temperature profile is now significantly hotter than the condensation curve on $MgSiO_3$ for the entire atmosphere and no clouds form. As illustrated by Figure 11, shifting the temperature-pressure profile will change the condensation pressure for the cloud model. The resulting albedo increases with cloud altitude up until the point where the atmosphere is too hot and the clouds would evaporate.

Each case from the parameter space shows a different peak, magnitude, and asymmetry when the full phase curve is plotted. To quantify the most plausible scenario from our chosen parameter space, the root-mean-square (RMS) error vs. the Demory et al. data was computed for each case and plotted in Figures 12 to 16. Figure 12 shows all the results for the non-temperature shifted models and Figures 13-16 separate the temperature-shifted models by cloud species. For a well-fit model, a clear, non-trivial, minimum will be present. The best-fit models are listed in Table 2. The standard deviation of the Kepler-7b observations is ~18.5 ppm with standard error $\sigma \sim 0.4$ ppm (*e.g.* fits above 19.3 ppm RMS error are outside the 95% confidence interval). Our RMS fit-metric approaches this number for our best-fit case (18.8 ppm) with only longitudinal variation and uniform clouds. To improve any further would require latitude-variation, different species, or a gradient in cloud thickness vs. longitude. Overall, $MgSiO_3$ models with a small $f_{sed}$ and a dayside cloud offset of 90° present the best-fit scenarios (RMS error within 1-$\sigma$ of the standard deviation of the data), and Fe dominated clouds should be ruled-out. Demory et al. suggest a bright hemisphere offset by $86 \pm 12°$. The phase curve maximum of the data is at $39.6 \pm 10.8°$. Figure 17 shows planet-star flux ratio of the best-fit model plotted over the Demory et al. data for the full range of cloud offsets. The brown curve in Figure 17 shows the 90° offset case. Figure 18 shows our predicted disk-integrated albedo spectrum for this case at a few points in the orbit.

4. CONCLUSION

Previous studies have shown that the tidally-locked conditions of a hot Jupiter create stark temperature contrasts between the dayside and nightside of the planet. The large thermal gradients will drive strong winds in these atmospheres that will in some cases cause global temperature distributions to be misaligned from what would have been expected based purely on incident energy considerations. This misalignment manifests as a shift in the maximum of the thermal emission phase curve of the planet. Here, we have shown that this temperature misalignment can produce a similar shift for reflected-light phase curves if offsets in the temperature map lead to partial (longitudinally-varying) cloud formation on the dayside of the planet. This study investigates how contributions from different albedo spectra at different geographic locations on the exoplanet would affect the planetary phase curve in the visible wavelengths (in this case, the differences in the spectra are due to clouds). The increased albedo from the presence of the clouds considered here greatly changes the visible light spectrum of the planet. We have considered two cases at a range of source-observer geometries (functions of the planet's orbital phase). We observed that the shift in the optical phase curve maximum increases as the cloud offset increases. For a hot Jupiter cloud offset of ~30-100°, observers may see phase curve shifts on the order of ~10-30° in phase angle (~0.05 fractional orbital phase). For cooler targets, like the Jupiter at 2.0 AU case, clouds may have a small effect on the alignment of the optical phase curve (~2-10°). Finally, we created a pipeline for computing and testing the albedo spectra of different cloud conditions against a phase curve data set. For the published Kepler-7b phase curve (Demory et al. 2013), this approach suggests a cloud offset of ~90° in the dayside hemisphere with $MgSiO_3$ clouds and a small sedimentation efficiency (i.e. thick, small particle clouds). Hypothesized Fe dominated clouds are ruled out by this method due to poor quality fits.

With planet-specific models to aid observation, reflected-light phase curves complement infrared observations in determining spatially resolved atmospheric properties. In future work, we will explore the extent to which this model can make phase curve predictions and interpret observations for exoplanets. We will investigate the effect of varying temperature-pressure profiles across the planet, cloud formations, atmospheric dynamics, and filter specifications. We will also explore how varying factors such as particle size, cloud deck thickness, altitude, and species can affect the shape of the phase curve. This approach can be extended to direct imaging candidates and can be combined with other reflected-light phase curve effects (e.g. rings, land/water mass, ellipsoidal variation), to characterize exoplanetary systems.

ACKNOWLEDGMENTS

The authors thank Brice-Olivier Demory for his helpful discussion and for providing the Kepler-7b phase curve data. This work was performed in part under contract with the Jet Propulsion Laboratory (JPL) funded by NASA through the Sagan Fellowship Program executed by the NASA Exoplanet Science Institute.

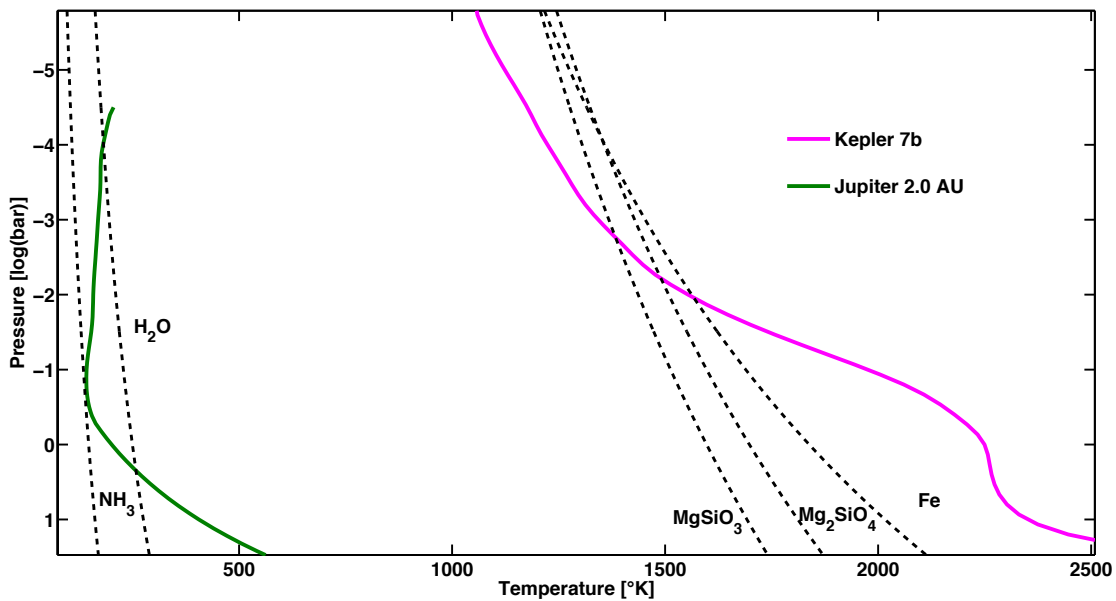

Figure 1: One-dimensional model pressure-temperature profiles derived from radiative-convective equilibrium simulations for a hypothetical Jupiter at 2.0 AU (green) and Kepler-7b (magenta), see Cahoy et al. 2010 and Demory et al. 2013. Condensation curves for ammonia ($NH_3$), water ($H_2O$), forsterite ($Mg_2SiO_4$), enstatite ($MgSiO_3$) and iron (Fe) are shown to give the approximate vertical

location of the cloud decks in each case. The albedo spectra model takes these pressure-temperature profiles as inputs to determine the opacity of an atmospheric layer.

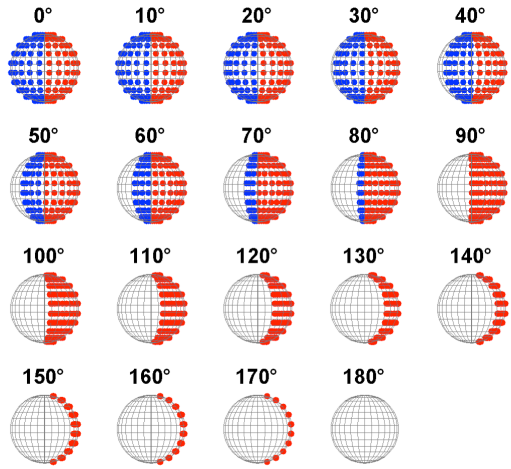

Figure 2: The disk integration and phase angle calculation uses Tchebyshev -Gauss integration similar to Cahoy et al. 2010, only with 100 Gaussian Angles and 10 Tchebyshev Angles. A selection of phase increments is shown here with the planet divided into cloudy (red) and cloudless (blue) hemispheres. Grey regions represent the nightside of the planet.

| **Cloud Offset** | **30°** | **90°** |
|---|---|---|
| Jupiter 2.0 AU | $2.0 \pm 0.4°$ | $10.1 \pm 0.8°$ |
| Kepler-7b | $9.8 \pm 1.7°$ | $17.8 \pm 2.0°$ |

Table 1: The shifted maxima of each phase curve due to the presence of offset clouds from the nightside of the planet into the dayside.

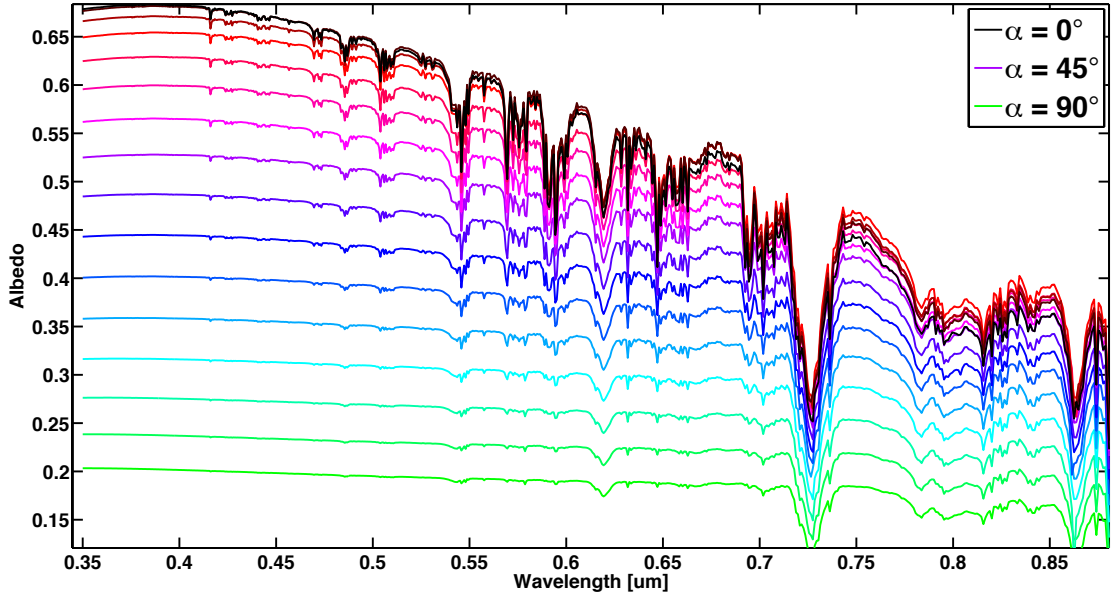

Figure 3: Albedo spectra of Jupiter at 2.0 AU with a 90° cloud offset for a range of phase angles from 0° (black) to 90° (green). The dominant absorber for the Jupiter model is $CH_4$. The first few initial $\alpha = 0°$-6° spectra (black) are clearly below the subsequent spectra (red-pink) at longer wavelengths. This leads to a phase curve maximum at $\alpha$ other than 0°.

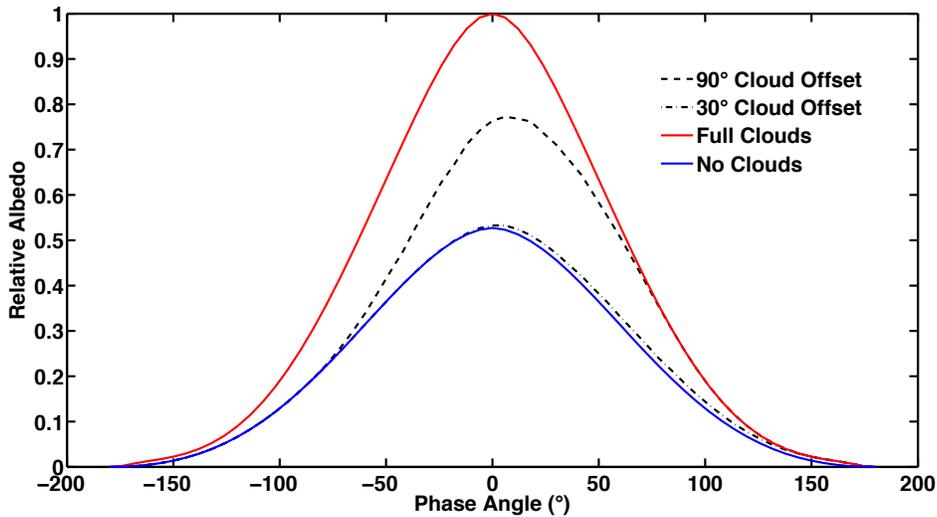

Figure 4: Reflected light phase curves of Jupiter at 2.0 AU for clear, cloudy, and two different shifted cloud coverage cases, 30° and 90°. Each phase curve is normalized to the full cloudy case. The phase curves that have non-uniform cloud coverage in the longitudinal direction (30°, 90° cloud offsets) have a maximum shifted away from secondary eclipse ($\alpha = 0°$).

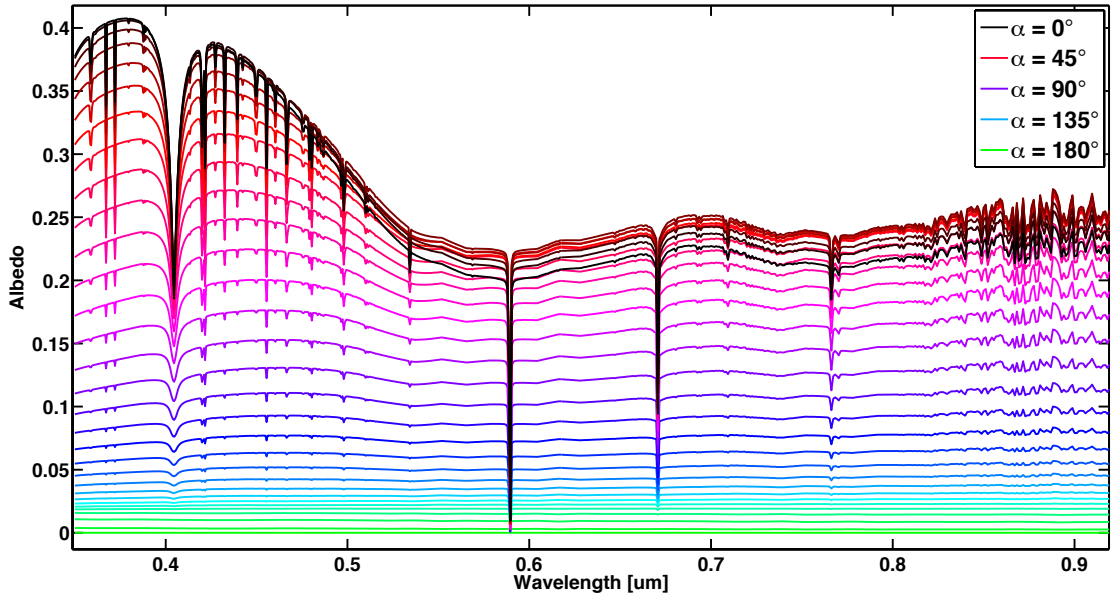

Figure 5: Albedo spectra of Kepler-7b with a 90° cloud offset for range of phase angles from 0° (black) to 180° (green). The dominant absorbers for Kepler-7b are neutral alkalis. The first few initial $\alpha = 0°$, $\alpha = 6°$, and $\alpha = 12°$ spectra (black) are clearly below the subsequent spectra (red-pink) at longer wavelengths. This leads to a phase curve maximum at $\alpha$ other than at 0°.

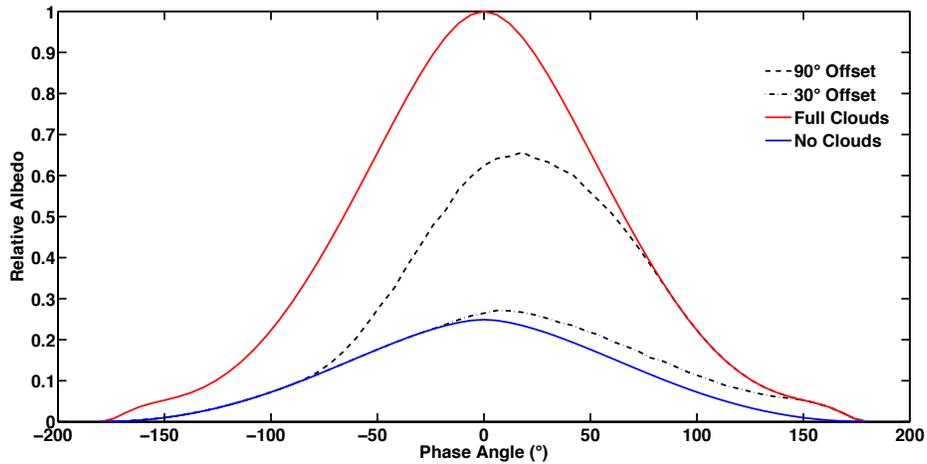

Figure 6: Reflected light phase curves of Kepler-7b for clear, cloudy, and 2 different shifted cloud coverage cases, 30° and 90°. Each phase curve is normalized to the cloudy case. The phase curves that have non-uniform cloud coverage in the longitudinal direction (30°, 90° cloud offsets) have a maximum shifted away from secondary eclipse ($\alpha = 0°$).

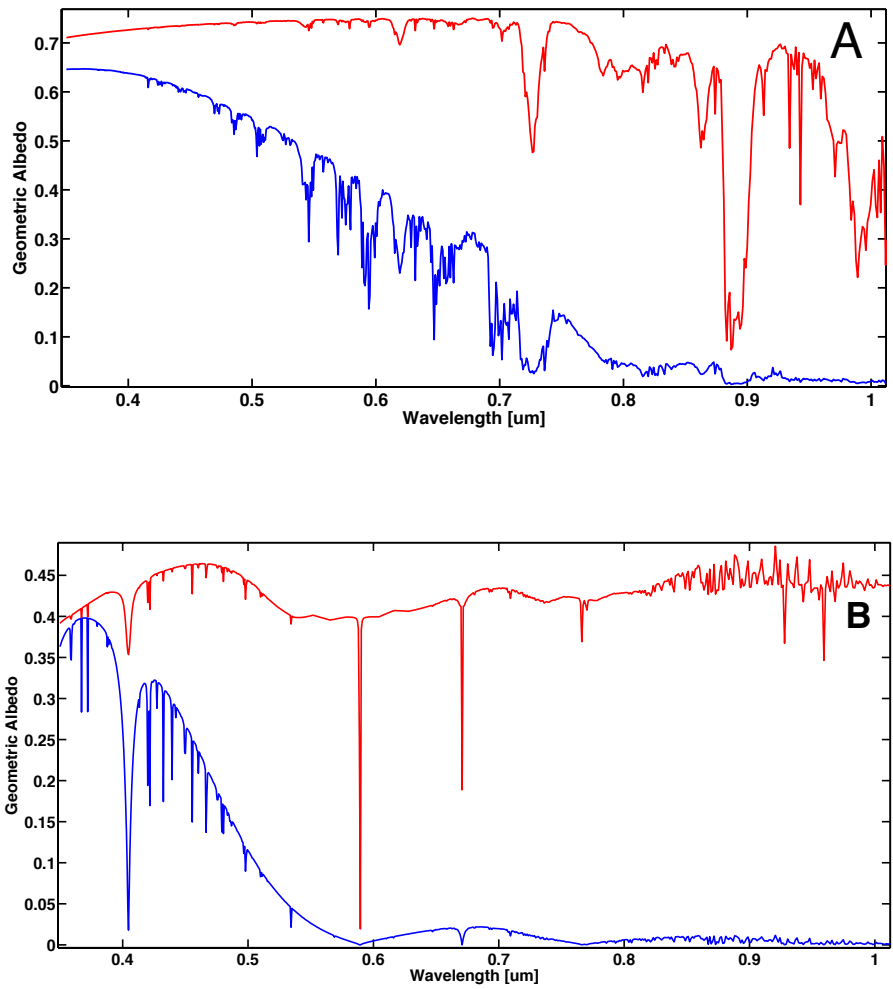

Figure 7: Albedo spectra of the full cloudy (red) and cloudless (blue) planets at $\alpha = 0°$. A) Jupiter at 2.0 AU; these spectra have the smallest overall difference of the two cases and therefore the smallest shift of the phase curve maxima. B) Kepler-7b has a larger difference between cloudy and cloudless spectra.

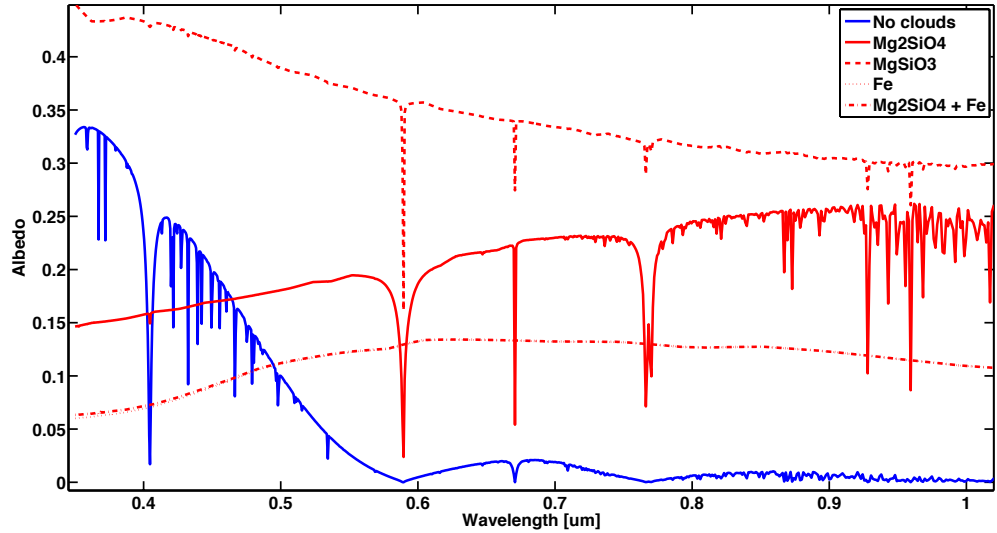

Figure 8: Variable Species, $f_{sed}$ = 0.1. The albedo spectra for varying cloud species are shown for a fixed $f_{sed}$ = 0.1. MgSiO$_3$ clouds show the most reflectivity for this condition.

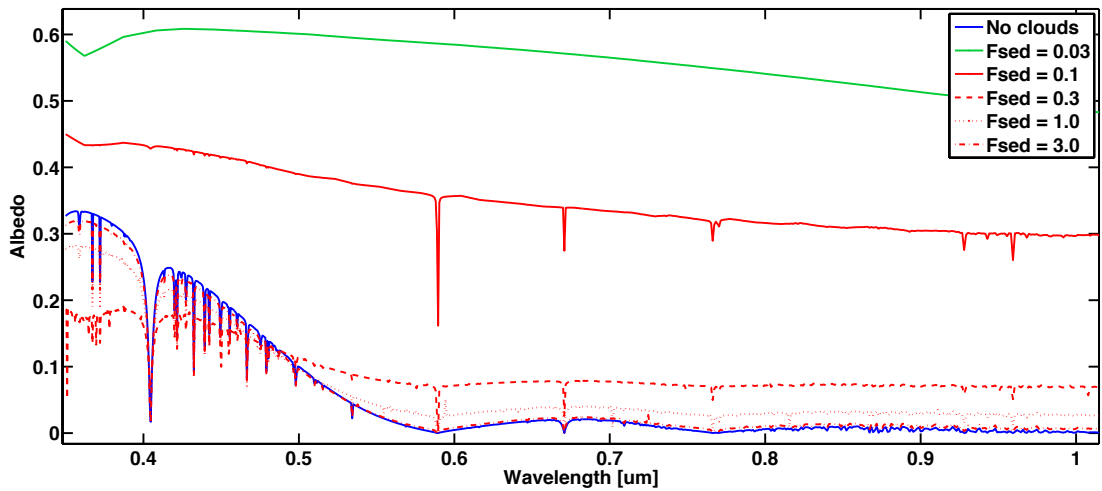

Figure 9: MgSiO$_3$ clouds, variable $f_{sed}$. The albedo spectra for MgSiO$_3$ clouds are shown at varying sedimentation efficiencies, $f_{sed}$. A lower $f_{sed}$ value suppresses sedimentation and increases the optical thickness of the cloud layer. As $f_{sed}$ increases, particle size increases. The effective particle sizes at the maximum concentration for each case are about 5, 15, 40, 100, and 220 μm respectively.

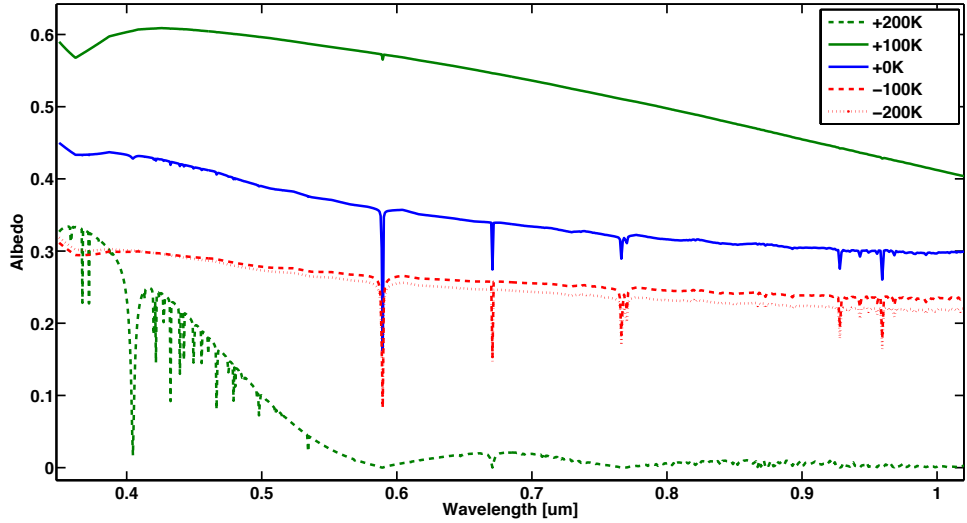

Figure 10: MgSiO$_3$ clouds, f$_{sed}$ = 0.1, shifted profiles. The pressure-temperature profile for Kepler-7b is shifted in temperature to simulate clouds formation at different altitudes (pressures). The resulting spectra are shown here for MgSiO$_3$, at $f_{sed}$ = 0.1. With a + 200 K shift, we see that clouds no longer form for this case.

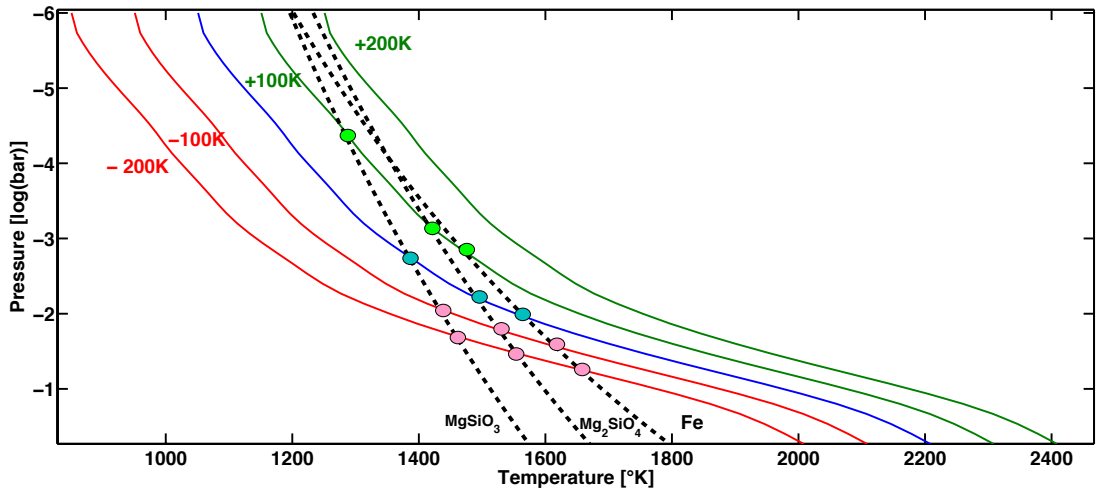

Figure 11: The intersections of the condensation curves with pressure-temperature profile determine the altitude of cloud formation. Shifting the pressure-temperature profile is used in this study to simulate the effect of varying cloud altitude on the albedo spectra.

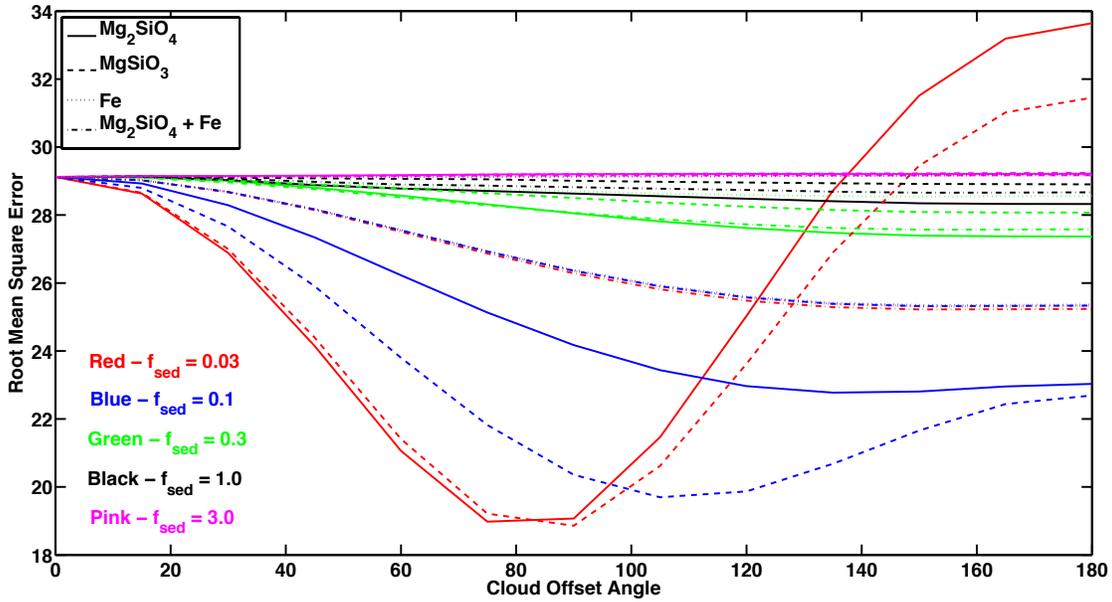

Figure 12: Goodness of Fit: Species vs. $f_{sed}$. The root mean square error (RMSE) is shown here vs. the cloud offset angle for the non-temperature shifted cases. The line pattern distinguishes the cloud species and color distinguishes the sedimentation efficiency. Minima in these curves represent well-fit cases.

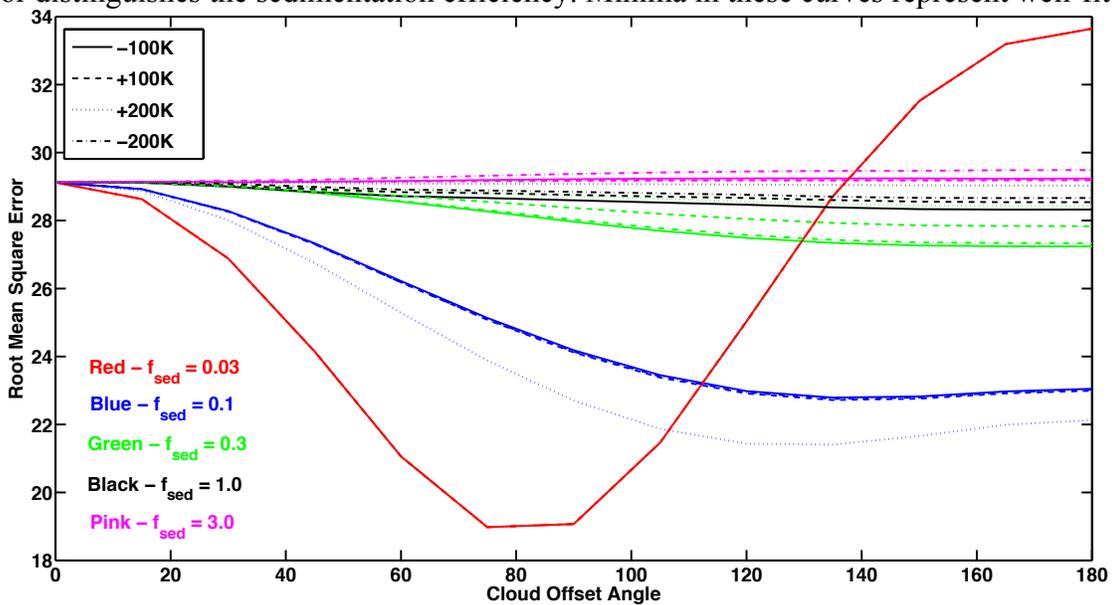

Figure 13: Goodness of Fit: $Mg_2SiO_4$ vs. $f_{sed}$ vs. temperature shift. The RMSE is shown here vs. the cloud offset angle for the temperature-shifted $Mg_2SiO_4$ cases. The line pattern distinguishes the temperature shift and color distinguishes the sedimentation efficiency. Minima in these curves represent well-fit cases.

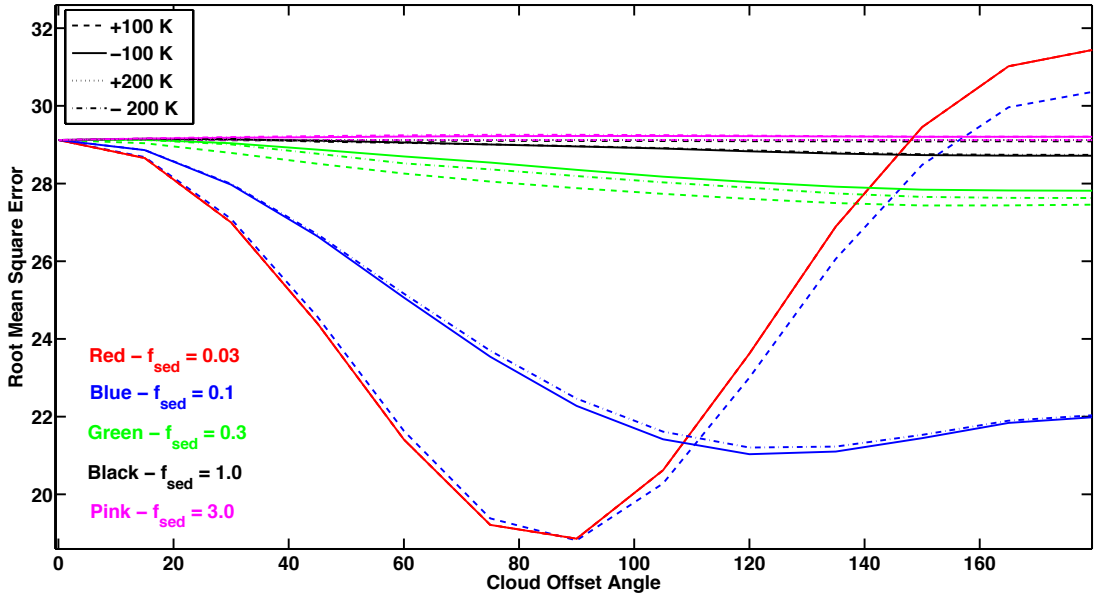

Figure 14: Goodness of Fit: $MgSiO_3$ vs. $f_{sed}$ vs. temperature shift. The RMSE is shown here vs. the cloud offset angle for the temperature-shifted $MgSiO_3$ cases. The line pattern distinguishes the temperature shift and color distinguishes the sedimentation efficiency. Minima in these curves represent well-fit cases.

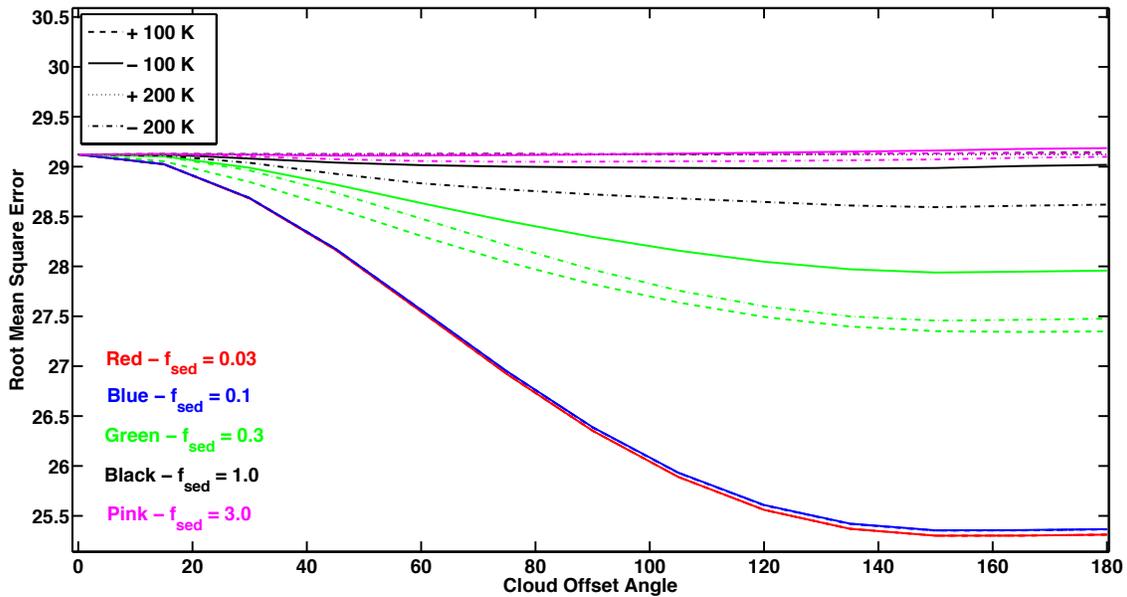

Figure 15: Goodness of Fit: Fe clouds vs. $f_{sed}$ vs. temperature shift. The RMSE is shown here vs. the cloud offset angle for the temperature-shifted Fe cases. The line pattern distinguishes the temperature shift and color distinguishes the sedimentation efficiency. Minima in these curves represent well-fit cases.

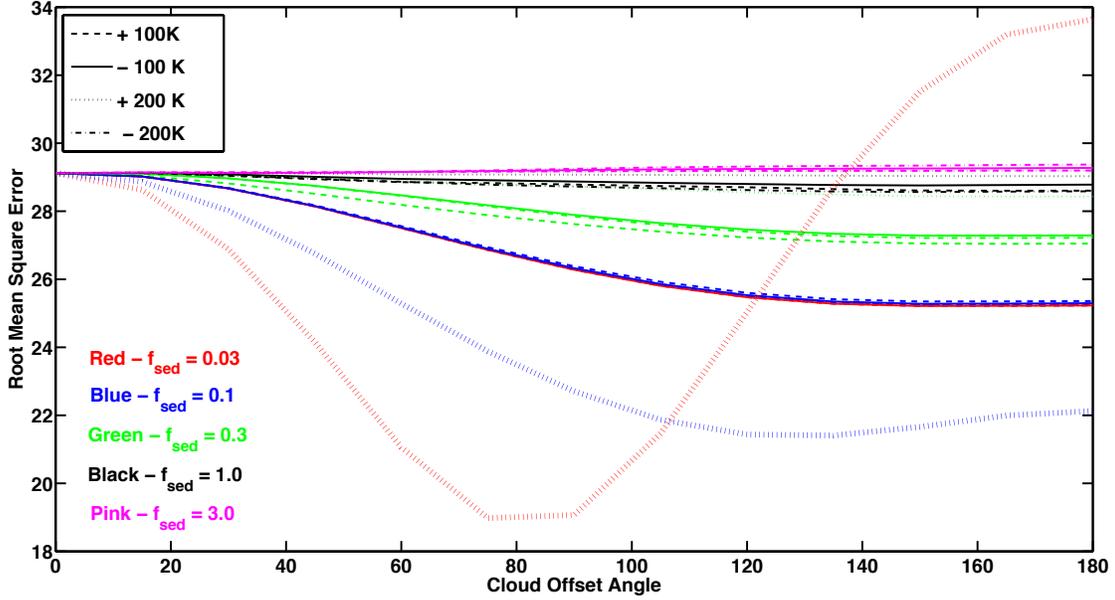

Figure: 16: Goodness of Fit: $Mg_2SiO_4$ + Fe vs. $f_{sed}$ vs. temperature shift. The RMSE is shown here vs. the cloud offset angle for the temperature-shifted $Mg_2SiO_4$ + Fe cases. The line pattern distinguishes the temperature shift and color distinguishes the sedimentation efficiency. Minima in these curves represent well-fit cases.

Table 2: Best-Fit Cloud Cases to Demory et al. (2013)

|  | Species | RMS minimum (ppm) | $f_{sed}$ | Cloud Offset | Temp. Shift |
|---|---|---|---|---|---|
|  | MgSiO3 | 18.8 | 0.1 | 90° | 100K |
|  | MgSiO3 | 18.9 | 0.03 | 90° | 0K |
|  | MgSiO3 | 18.9 | 0.03 | 90° | -100K |
| 1σ | Mg2SiO4 | 19.0 | 0.03 | 75° | 0K |
|  | Mg2SiO4 | 19.0 | 0.03 | 75° | -100K |
| 2σ | Mg2SiO4 + Fe | 19.4 | 0.03 | 75° | 200K |
|  | MgSiO3 | 19.7 | 0.1 | 105° | 0K |
| 3σ | MgSiO3 | 21.0 | 0.1 | 120° | -100K |
|  | MgSiO3 | 21.2 | 0.1 | 120° | -100K |
|  | Mg2SiO4 | 21.4 | 0.1 | 135° | 200K |
|  | Mg2SiO4 + Fe | 21.4 | 0.1 | 135° | 200K |
|  | Mg2SiO4 | 22.7 | 0.1 | 135° | 100K |
|  | Mg2SiO4 | 22.8 | 0.1 | 135° | -100K |
|  | Mg2SiO4 | 22.8 | 0.1 | 135° | 0K |
|  | Mg2SiO4 | 22.8 | 0.1 | 135° | -200K |
|  | Fe | 25.2 | 0.03 | 150° | 0K |

Table 2: A list of the 16 best-fit cloud cases (from 1300 cases) to the Demory et al. (2013) phase curve. Cases are ranked by their RMS error. Silicate clouds, with a low $f_{sed}$, are clearly favored with a cloud offset near 90°. The dotted lines denoted the 1-σ, 2-σ, and 3-σ, difference between for the RMS error and the standard deviation of the observation.

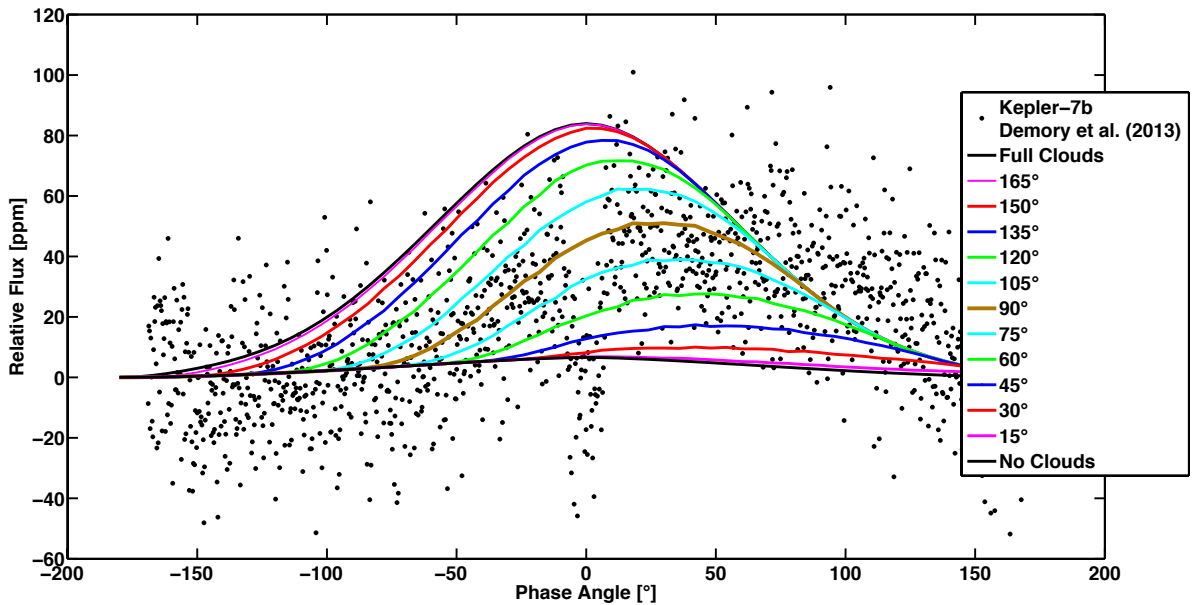

Figure 17: The best-fit cloud condition for a range of offset angles is plotted over the planet-star flux ratio measured by Demory et al. (2013). The brown curve shows the overall best-fit (MgSiO$_3$, $f_{sed}$ = 0.1, 90° offset). Our albedo spectra are multiplied by the stellar spectrum, along with the Kepler transmission function, then scaled by the square of the ratio between the planet radius and semi-major axis.

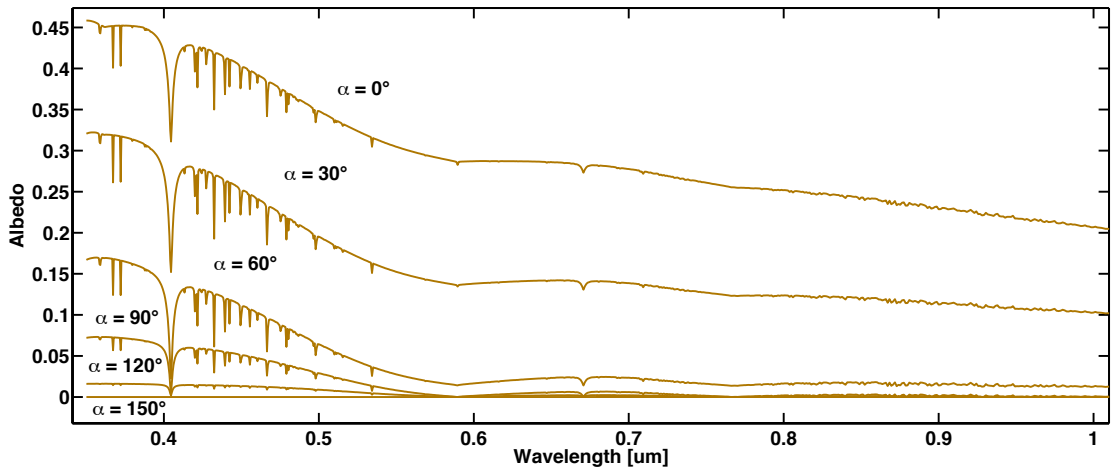

Figure 18: The disk-integrated, best-fit Kepler-7b albedo spectrum (MgSiO$_3$, $f_{sed}$ = 0.1) is shown for a range of orbital phase angles.